\documentclass[12pt]{article}
\usepackage{epsf}
\begin{document}

\title{Spectral Density of Sparse \\ Sample Covariance Matrices} 

\author{Taro Nagao$^*$ and Toshiyuki Tanaka$^\dagger$}  
 
\date{}
\maketitle

\begin{center}
\it
$^*$ Graduate School of Mathematics,
Nagoya University, Chikusa-ku, \\ Nagoya 464-8602, Japan \\ 
\it
$^\dagger$ Graduate School of Informatics, Kyoto University, 
Sakyo-ku, \\ Kyoto 606-8501, Japan \\
\end{center} 

\begin{abstract}

Applying the replica method of statistical mechanics, we 
evaluate the eigenvalue density of the large random matrix 
(sample covariance matrix) of the form $J = A^{\rm T} A$, 
where $A$ is an $M \times N$ real sparse random matrix. 
The difference from a dense random matrix is the most 
significant in the tail region of the spectrum. We compare 
the results of several approximation schemes, focusing 
on the behavior in the tail region. 

\end{abstract}

PACS: 02.50.-r;02.10.Yn

\medskip

Keywords: replica method; sample covariance matrices; random matrices

\medskip

\section{Introduction}
\setcounter{equation}{0}
\renewcommand{\theequation}{1.\arabic{equation}} 

Investigation on sample covariance matrices has 
a long history. It was originally introduced in 
multivariate statistical analysis~\cite{JW}. Let us 
consider $N$ dimensional sample vectors $X^{(m)}$, 
$m = 1,2,\cdots,L$, where $L$ is the number of samples. 
The sample covariance matrix $S$ is then defined as
\begin{equation}
S_{jl} = \frac{1}{L-1} \sum_{m=1}^L (X^{(m)}_j - 
{\bar X}_j) (X^{(m)}_l - {\bar X}_l),
\end{equation}
where the sample mean vector ${\bar X}$ is 
\begin{equation}
{\bar X}= \frac{1}{L} \sum_{m=1}^L X^{(m)}.
\end{equation} 
\par 
Let us suppose that $X^{(m)}$ are random 
vectors: the components $X_j^{(m)}$, $j = 
1,\,2,\,\cdots,\,N$, are independent and identically 
distributed Gaussian random variables. Then the 
distribution of $(L-1) S$ is the same as the 
distribution of the $N \times N$ matrix 
\begin{equation}
J = A^{\rm T} A  
\end{equation}
($A^{\rm T}$ is the transpose of $A$), where 
the elements of the $(L-1) \times N$ matrix $A$ are 
independent and identically distributed 
Gaussian random variables. The ensemble of 
sample covariance matrices is sometimes 
called ``chiral Gaussian'' or ``Laguerre'' 
ensemble and finds applications in neural-network 
learning~\cite{CKS}, quantum chromodynamics~\cite{JV}, 
mesoscopic physics~\cite{CB}, finance~\cite{LCBP,PLEROU} 
and wireless communication~\cite{TV}. 
\par
In multivariate statistics one is usually 
interested in the limit $L \rightarrow 
\infty$ with fixed $N$. On the other hand, 
motivated by Wigner's 
celebrated work~\cite{WIG} on real symmetric 
random matrices, Mar\u{c}enko and Pastur 
studied another limit $N,L \rightarrow \infty$ with 
$L/N \rightarrow \alpha$ ($0 < \alpha \leq 1$) and 
derived the asymptotic density for the 
scaled eigenvalues of $J$ as~\cite{MP,PASTUR}
\begin{equation}
\rho(x) = (1 - \alpha) \delta(x) + 
\frac{1}{2 \pi x} \sqrt{(x - (\sqrt{\alpha}-1)^2)(
(\sqrt{\alpha} + 1)^2 - x)}.
\end{equation}
This asymptotic result is valid 
for more general distributions of the 
matrix $A$, and is sometimes called 
Mar\u{c}enko-Pastur law. 
\par
One of the simplest ways to modify the random 
matrix $J$ so that Mar\u{c}enko-Pastur law breaks 
down is to make the matrix $A$ sparse. An example demonstrating 
the significance of considering random matrix ensembles defined 
on the basis of sparse $A$ can be found in communication theory: 
Information-theoretic channel capacity of a randomly-spread 
Code-Division Multiple-Access (CDMA) channel is evaluated in 
terms of eigenvalue distribution of the matrix $J=A^{\rm T}A$ 
with $A$ defining the random spreading. It has been argued~\cite{YT} 
that some wideband CDMA schemes can be modeled as a sparsely-spread 
system, where deviations from Mar\u{c}enko-Pastur law may affect 
performance of such systems. Sparse random matrices in general are 
also of interest in many branches of applications. In particular, 
the eigenvector localization expected to occur in sparse random 
matrices is one of the most outstanding phenomena  
in disordered systems. The appearance of 
isolated eigenvalue spectra in the tail region is 
another interesting feature. Such features are 
also observed in heavy-tailed random matrices 
and were recently studied in detail~\cite{SOSH,SF, BBP}. 
\par
In this paper, as a continuation of a brief report 
by one of the authors~\cite{TT}, we investigate the 
spectral (eigenvalue) density of sparse sample 
covariance matrices.  In the case of real symmetric sparse 
random matrices $R$ whose elements $R_{jl}$ ($j \leq l$) 
are independently distributed, the asymptotic spectral 
density was already studied by several authors~\cite{RB,RD,MF,SC}. 
Field theoretic (replica and supersymmetry) methods 
were usually employed. For example, Semerjian and 
Cugliandolo utilized a sophisticated version of 
replica method and discussed the spectral density 
in connection with the percolation problem~\cite{SC}. 
In the next section, we present a similar replica 
method developed for the analysis of sparse sample 
covariance matrices. In \S 3 and \S 4, the effective 
medium approximation (EMA) and its symmetrized version 
are introduced. In \S 5, we consider the case 
of binary distribution and evaluate the 
asymptotic spectral density. In \S 6, the tail 
behavior of the spectral density is analyzed 
by means of the single defect approximation (SDA), 
which improves the symmetrized version of EMA.  

\section{Replica Method}
\setcounter{equation}{0}
\renewcommand{\theequation}{2.\arabic{equation}}

We are interested in the asymptotic spectral density 
of the $N \times N$ matrix 
\begin{equation}
J = A^{\rm T} A, 
\end{equation}
where $A$ is an $M \times N$ real sparse random matrix($M \leq N$). 
The elements of $A$ are independently distributed according to 
the probability distribution function
\begin{equation}
P(x) = \left( 1 - \frac{p}{N} \right) \delta(x) + \frac{p}{N} 
\Pi(x) 
\end{equation} 
with a fixed positive number $p < N$. Here $\delta(x)$ is 
Dirac's delta function. The asymptotic limit 
$N \rightarrow \infty$ with $p$ and $\alpha = M/N$ fixed 
will be considered. We assume that $\Pi(x)$ does not depend 
on $N$ and has zero measure on the origin:
\begin{equation}
\lim_{\epsilon \searrow 0} \int_{-\epsilon}^{\epsilon} 
\Pi(x)\, {\rm d}x = 0. 
\end{equation}
\par 
The Fourier transform of $P(x)$ is given by
\begin{equation}
\label{fourier}
{\hat P}(k) =  \int_{-\infty}^{\infty} P(x) 
\ {\rm e}^{- i k x}\, {\rm d}x  =  1 + \frac{p}{N} f(k), 
\end{equation}  
where
\begin{equation}
f(k) =  \int_{-\infty}^{\infty} \Pi(x) \ {\rm e}^{- i k x}\, {\rm d}x - 1. 
\end{equation} 
\par
Let us denote the average over $MN$ copies of $P(x)$ by 
brackets $\langle\cdot\rangle$. Then the average spectral 
density of $J$ is defined as 
\begin{equation}
\rho(\mu) = \left\langle \frac{1}{N} \sum_{j=1}^N \delta(\mu - \mu_j) 
\right\rangle,
\end{equation}
where $\mu_1,\,\mu_2,\,\cdots,\,\mu_N$ are the eigenvalues of $J$. 
We can rewrite $\rho(\mu)$ as
\begin{eqnarray}
\rho(\mu) & = & \frac{1}{N \pi} {\rm Im} {\rm Tr} 
\left\langle (J - (\mu + i \epsilon) I)^{-1} \right\rangle 
\nonumber \\ & = & 
\frac{2}{N \pi} {\rm Im}  
\frac{\partial}{\partial \mu} \left\langle 
\ln Z(\mu + i \epsilon) \right\rangle,
\end{eqnarray}
where $I$ is an $N \times N$ identity matrix, 
$\epsilon$ is a positive infinitesimal 
number and   
\begin{equation}
Z(\mu^{\prime}) = \int \prod_{j=1}^N {\rm d}\phi_j 
\ {\rm exp}\left\{ 
\frac{i \mu^{\prime}}{2} \sum_{j=1}^N \phi_j^2 - \frac{i}{2} 
\sum_{j=1}^N \sum_{l=1}^N 
J_{jl} \phi_j \phi_l \right\}, \ \ \ \mu^{\prime} = \mu + i \epsilon 
\end{equation}
is called the partition function. In order to 
evaluate the average over $P(x)$, using the 
relation
\begin{equation}
\lim_{n \rightarrow 0} \frac{\partial}{\partial n}\ln\langle Z^n\rangle = \langle\ln Z\rangle,
\end{equation} 
we deduce 
\begin{eqnarray}
\rho(\mu) & = & 
\frac{2}{N \pi} {\rm Im}  
\frac{\partial}{\partial \mu}\left\{\lim_{n\to0}\frac{\partial}{\partial n}\ln\left\langle 
(Z(\mu + i \epsilon))^n \right\rangle\right\} \nonumber \\  
& = & \lim_{n \rightarrow 0}
\frac{2}{\pi n} {\rm Im}  
\frac{\partial}{\partial \mu} \frac{1}{N}\ln \left\langle 
(Z(\mu + i \epsilon))^n \right\rangle. 
\end{eqnarray}
Thus we find that the spectral density can be written 
in terms of the average $\langle Z^n \rangle$.  
\par 
Noting
\begin{equation}
\int \prod_{k=1}^M {\rm d}\psi_k \ {\rm exp}\left\{  \frac{i}{2} 
\sum_{k=1}^M \left( \psi_k - \sum_{j=1}^N \phi_j A_{kj} \right) 
\left( \psi_k - \sum_{l=1}^N \phi_l A_{kl} \right) \right\}= 
(2 \pi i)^{M/2},
\end{equation}
we find 
\begin{eqnarray}
& & \frac{1}{(2 \pi i)^{M/2}} \int \prod_{k=1}^M {\rm d}\psi_k 
\ {\rm exp}\left\{ \frac{i}{2} \sum_{k=1}^M \psi_k^2 - i \sum_{k=1}^M 
\sum_{j=1}^N \psi_k A_{kj} \phi_j \right\} \nonumber \\ & = & 
{\rm exp}\left\{ - \frac{i}{2} \sum_{k=1}^M \sum_{j=1}^N 
\sum_{l=1}^N \phi_j A_{kj} \phi_l A_{kl} \right\} 
= {\rm exp} \left\{ - \frac{i}{2} \sum_{j=1}^N \sum_{l=1}^N 
J_{jl} \phi_j \phi_l \right\}, \nonumber \\ 
\end{eqnarray}
from which it follows that
\begin{eqnarray}
Z(\mu^{\prime}) & = & \frac{1}{(2 \pi i)^{M/2}} 
\int \prod_{j=1}^N {\rm d}\phi_j 
\prod_{k=1}^M {\rm d}\psi_k 
\ {\rm exp}\left\{ 
\frac{i \mu^{\prime}}{2} \sum_{j=1}^N \phi_j^2 + 
\frac{i}{2} \sum_{k=1}^M \psi_k^2 \right\} 
\nonumber \\ 
& \times & {\rm exp}\left\{  
- i  \sum_{k=1}^M \sum_{j=1}^N 
\psi_k A_{kj} \phi_j \right\}.
\end{eqnarray}
Now we introduce the vectors (replica variables)
\begin{equation}
{\vec \phi}_j = 
(\phi^{(1)}_j,\phi^{(2)}_j,\cdots,\phi^{(n)}_j), \ \ \ 
{\vec \psi}_k =  
(\psi^{(1)}_k,\psi^{(2)}_k,\cdots,\psi^{(n)}_k)
\end{equation}
and the corresponding measures
\begin{equation}
{\rm d}{\vec \phi}_j =  
{\rm d}\phi^{(1)}_j {\rm d}\phi^{(2)}_j \cdots {\rm d} \phi^{(n)}_j, \ \ \  
{\rm d}{\vec \psi}_k =  
{\rm d}\psi^{(1)}_k {\rm d} \psi^{(2)}_k\cdots {\rm d} \psi^{(n)}_k 
\end{equation}
so that the average $\langle Z^n \rangle$ can 
be rewritten as
\begin{eqnarray}
\langle Z^n \rangle & = & \frac{1}{(2 \pi i)^{Mn/2}} 
\int \prod_{j=1}^N {\rm d}{\vec \phi}_j 
\prod_{k=1}^M {\rm d}{\vec \psi}_k 
\ {\rm exp}\left\{ 
\frac{i \mu^{\prime}}{2} \sum_{j=1}^N {\vec \phi}_j^2 + 
\frac{i}{2} \sum_{k=1}^M {\vec \psi}_k^2 \right\} \nonumber \\  
& \times & \left\langle {\rm exp}\left\{ 
- i  \sum_{k=1}^M \sum_{j=1}^N 
A_{kj} {\vec \psi}_k \cdot 
{\vec \phi}_j \right\} \right\rangle,
\end{eqnarray}
where ${\vec \phi}_j^2 = {\vec \phi}_j \cdot {\vec \phi}_j$ and 
${\vec \psi}_k^2 = {\vec \psi}_k \cdot {\vec \psi}_k$.  
\par
Using the Fourier transform (\ref{fourier}) of the probability 
distribution function $P(x)$, we obtain
\begin{eqnarray}
\left\langle {\rm exp}\left\{ 
- i  \sum_{k=1}^M \sum_{j=1}^N 
A_{kj} {\vec \psi}_k \cdot 
{\vec \phi}_j \right\} \right\rangle & = & 
\prod_{k=1}^M \prod_{j=1}^N 
\left( \int {\rm d}x\, P(x) \ {\rm exp}\left\{ - i 
{\vec \psi}_k \cdot {\vec \phi}_j 
x \right\} \right) \nonumber \\ 
& \sim &  \prod_{k=1}^M \prod_{j=1}^N {\rm exp}\left\{ 
\frac{p}{N} f({\vec \psi}_k \cdot {\vec \phi}_j) \right\}
\end{eqnarray}
in the limit $N \rightarrow \infty$. It follows that 
\begin{eqnarray}
\label{zn1}
\langle Z^n \rangle & \sim & \frac{1}{(2 \pi i)^{Mn/2}} 
\int \prod_{j=1}^N {\rm d}{\vec \phi}_j 
\prod_{k=1}^M {\rm d}{\vec \psi}_k 
\ {\rm exp}\left\{ 
\frac{i \mu^{\prime}}{2} \sum_{j=1}^N {\vec \phi}_j^2 + 
\frac{i}{2} \sum_{k=1}^M {\vec \psi}_k^2 \right\} \nonumber \\  
& \times & {\rm exp}\left\{ 
\frac{p}{N} \sum_{k=1}^M \sum_{j=1}^N 
f({\vec \psi}_k \cdot {\vec \phi}_j) \right\}.
\end{eqnarray}

\section{Effective Medium Approximation}
\setcounter{equation}{0}
\renewcommand{\theequation}{3.\arabic{equation}}

We are in a position to explain how to 
evaluate the spectral density by means of a 
scheme called the effective medium approximation 
(EMA)~\cite{SC}. We first rewrite (\ref{zn1}) as 
\begin{eqnarray}
\langle Z^n \rangle  
& \sim & 
\int \prod_{j=1}^N {\rm d}{\vec \phi}_j 
\ {\rm exp}\left\{ \frac{i \mu^{\prime}}{2} 
\sum_{j=1}^N {\vec \phi}_j^2 \right\} 
\nonumber \\ & \times & 
\left[ \frac{1}{(2 \pi i)^{n/2}} 
\int {\rm d}{\vec \psi} 
\ {\rm exp} \left\{ 
\frac{i}{2} {\vec \psi}^2   
+ \frac{p}{N} \sum_{j=1}^N 
f({\vec \psi} \cdot {\vec \phi}_j) \right\} \right]^M. 
\end{eqnarray}
Introducing 
\begin{equation}
{\tilde \xi}({\vec \phi})  = \frac{1}{N} \sum_{j=1}^N \delta({\vec \phi} - {\vec \phi}_j),
\end{equation}
we find
\begin{eqnarray}
\langle Z^n \rangle  
& \sim & 
\int \prod_{j=1}^N {\rm d}{\vec \phi}_j 
\ {\rm exp}\left\{ N \frac{i \mu^{\prime}}{2} 
\int {\rm d}{\vec \phi}\, {\tilde \xi}({\vec \phi}) {\vec \phi}^2 \right\} 
\nonumber \\ & \times & 
\left[ \frac{1}{(2 \pi i)^{n/2}} 
\int {\rm d}{\vec \psi} 
\ {\rm exp} \left\{ 
\frac{i}{2} {\vec \psi}^2   
+ p \int {\rm d}{\vec \phi}\, {\tilde \xi}({\vec \phi}) 
f({\vec \psi} \cdot {\vec \phi}) \right\} \right]^M. 
\end{eqnarray}
\par
Let us introduce an order parameter function $\xi({\vec \phi})$, which is 
normalized as
\begin{equation}
\int \xi({\vec \phi})\,{\rm d}{\vec \phi} = 1.
\end{equation}
Then, using the functional integral of the delta function
\begin{equation}
\int {\cal D}\xi({\vec \phi}) \prod_{{\vec \phi}} \delta(\xi({\vec \phi}) - 
{\tilde \xi}({\vec \phi})) = 1, 
\end{equation}
we obtain
\begin{equation}
\label{zn2}
\langle Z^n \rangle \sim   
\int {\cal D}\xi({\vec \phi}) \int \prod_{j=1}^N {\rm d}{\vec \phi}_j 
\prod_{{\vec \phi}} \delta(\xi({\vec \phi}) - 
{\tilde \xi}({\vec \phi}))  Q[\xi({\vec \phi})],
\end{equation} 
where
\begin{eqnarray}
Q[\xi({\vec \phi})] & = & {\rm exp}\left\{ N \frac{i \mu^{\prime}}{2} 
\int {\rm d}{\vec \phi}\, \xi({\vec \phi}) {\vec \phi}^2 \right\} 
\nonumber \\ & \times & 
\left[ \frac{1}{(2 \pi i)^{n/2}} 
\int {\rm d}{\vec \psi} 
\ {\rm exp} \left\{ 
\frac{i}{2} {\vec \psi}^2   
+ p \int {\rm d}{\vec \phi}\, \xi({\vec \phi}) 
f({\vec \psi} \cdot {\vec \phi}) \right\} \right]^M. 
\end{eqnarray}
Moreover, we can utilize the Fourier transform of 
the delta function
\begin{equation}
\prod_{{\vec \phi}} \delta(\xi({\vec \phi}) - 
{\tilde \xi}({\vec \phi})) = \int {\cal D}c({\vec \phi}) 
\ {\rm exp}\left\{2 \pi i \int {\rm d}{\vec \phi}\, c({\vec \phi}) (\xi({\vec \phi}) 
- {\tilde \xi}({\vec \phi})) \right\}
\end{equation}  
in order to rewrite (\ref{zn2}) as
\begin{equation}
\langle Z^n \rangle \sim   
\int {\cal D}\xi({\vec \phi}) 
\int {\cal D}c({\vec \phi}) 
\ {\rm exp}\left\{ 
2 \pi i \int {\rm d}{\vec \phi}\, c({\vec \phi}) 
\xi({\vec \phi}) - N F[c({\vec \phi})] \right\} Q[\xi({\vec \phi})],
\end{equation} 
where
\begin{eqnarray}
F[c({\vec \phi})] & = & - \frac{1}{N} \ln 
\int \prod_{j=1}^N {\rm d}{\vec \phi}_j \ {\rm exp}\left\{ 
- 2 \pi i \int {\rm d}{\vec \phi}\, c({\vec \phi}) 
{\tilde \xi}({\vec \phi}) \right\} \nonumber \\ 
& = & - \frac{1}{N} \ln 
\int \prod_{j=1}^N {\rm d}{\vec \phi}_j \ {\rm exp}\left\{ 
- \frac{2 \pi i}{N} \int {\rm d}{\vec \phi}\, c({\vec \phi}) 
\sum_{j=1}^N \delta( {\vec \phi} - {\vec \phi}_j) 
\right\} \nonumber \\ 
& = & - \ln 
\int {\rm d}{\vec \varphi} \ {\rm exp}\left\{ 
- \frac{2 \pi i}{N} \int {\rm d}{\vec \phi}\, c({\vec \phi}) 
\delta( {\vec \phi} - {\vec \varphi}) 
\right\}. 
\end{eqnarray} 
\par
The major contribution to the functional integral over 
$c({\vec \phi})$ in the limit $N \rightarrow \infty$ 
comes from the stationary point $c^*({\vec \phi})$ 
satisfying
\begin{equation}
\left. \frac{\delta}{\delta c} \left( 
2 \pi i \int {\rm d}{\vec \phi}\, c({\vec \phi}) 
\xi({\vec \phi}) - N F[c({\vec \phi})] \right) \right|_{c = c^*} 
= 2 \pi i (\xi({\vec \phi}) - {\rm e}^{F - (2 \pi i/N) c^*({\vec \phi})}) = 0.
\end{equation}   
Hence we arrive at
\begin{eqnarray}
\label{zn3}
\langle Z^n \rangle & \sim &   
\int {\cal D}\xi({\vec \phi}) 
\ {\rm exp}\left\{ 
2 \pi i \int {\rm d}{\vec \phi}\, c^*({\vec \phi}) 
\xi({\vec \phi}) - N F[c^*({\vec \phi})] \right\} Q[\xi({\vec \phi})] \nonumber \\ 
& = & \int {\cal D}\xi({\vec \phi}) \ {\rm exp}\left\{ 
- N \int {\rm d}{\vec \phi}\, \xi({\vec \phi}) \ln \xi({\vec \phi})  
\right\} 
Q[\xi({\vec \phi})]. 
\end{eqnarray}
Let us rewrite (\ref{zn3}) as 
\begin{equation}
\langle Z^n \rangle \sim \frac{1}{(2 \pi i)^{Mn/2}} 
\int {\cal D}\xi({\vec \phi}) \ {\rm e}^{N S[\xi]},
\end{equation}
where
\begin{eqnarray}
S[\xi] & = & - \int {\rm d}{\vec \phi}\, \xi({\vec \phi}) \ln \xi({\vec \phi}) 
+ \frac{i \mu^{\prime}}{2} \int {\rm d}{\vec \phi}\, \xi({\vec \phi}) 
{\vec \phi}^2 \nonumber \\ 
& + & \alpha \ln \left[ \int {\rm d}{\vec \psi} 
\ {\rm exp}\left\{ \frac{i}{2} {\vec \psi}^2 + p \int {\rm d}{\vec \phi}\, 
\xi({\vec \phi}) f({\vec \psi} \cdot {\vec \phi}) \right\} \right] 
\end{eqnarray}
with $\alpha = M/N$.
\par
In the limit $N \rightarrow \infty$, the stationary point determined 
by the equation
\begin{equation}
\frac{\delta S[\xi]}{\delta \xi({\vec \phi})} = 0 
\end{equation}
gives the major contribution. However it is hard to solve it 
directly. Instead, we introduce a Gaussian ansatz
\begin{equation}
\xi^{\rm EMA}({\vec \phi}) = \frac{1}{(2 \pi i \sigma)^{n/2}} \ {\rm exp}\left\{ 
- \frac{{\vec \phi}^2}{2 i \sigma} \right\}, \ \ \ 
{\rm Im}\sigma(\mu^{\prime}) \leq 0,
\end{equation}
with a parameter $\sigma$, and look for a solution 
of the modified problem
\begin{equation}
\label{vema}
\frac{\partial S[\xi^{\rm EMA}]}{\partial \sigma} = 0.  
\end{equation}
This scheme is called the Effective Medium Approximation (EMA)~\cite{SC}. 
The condition ${\rm Im}\sigma(\mu^{\prime}) \leq 0$ is necessary 
to ensure the convergence of the integral over ${\vec \phi}$. 
\par
Putting the Gaussian ansatz into (\ref{vema}) and taking the 
limit $n \rightarrow 0$, we find
\begin{equation}
\label{ema}
1 - \sigma \mu - \alpha + \alpha \sum_{k=0}^{\infty} 
\frac{{\rm e}^{-p} p^k}{k!} \int 
\prod_{j=1}^k \left( 
\Pi(x_j) {\rm d}x_j \right) 
\frac{1}{1 - \sigma \sum_{j=1}^k x_j^2} = 0.
\end{equation}
Let us suppose that $\sigma$ satisfies this equation. 
Then the spectral density is evaluated as
\begin{eqnarray}
\label{emarho}
\rho(\mu) & = & \lim_{n \rightarrow 0}
\frac{2}{\pi n} {\rm Im} 
\frac{\partial}{\partial \mu} \frac{1}{N}\ln \left\langle 
(Z(\mu))^n \right\rangle \nonumber \\ 
& \sim & \lim_{n \rightarrow 0}
\frac{2}{\pi n} {\rm Im}  
\frac{\partial}{\partial \mu} S[\xi^{\rm EMA}] 
\nonumber \\ 
& = & \lim_{n \rightarrow 0}
\frac{1}{\pi n} {\rm Re} \int {\rm d}{\vec \phi}\,\xi({\vec \phi}) 
{\vec \phi}^2 \nonumber \\ 
& = & - \frac{1}{\pi} {\rm Im}\sigma.    
\end{eqnarray}
The spectral density can be calculated by 
numerically solving the stationary point 
equation (\ref{ema}) and by putting the 
solution into (\ref{emarho}).  
\par
We remark that there is another approximation scheme similar to EMA, 
which we call the Dual Effective Medium Approximation (DEMA).
Rewriting (\ref{zn1}) as   
\begin{eqnarray}
\langle Z^n \rangle  
& \sim & 
\int \prod_{k=1}^M {\rm d}{\vec \psi}_k 
\ {\rm exp}\left\{ \frac{i}{2} 
\sum_{k=1}^M {\vec \psi}_k^2 \right\} 
\nonumber \\ & \times & 
\left[ \frac{1}{(2 \pi i)^{\alpha n/2}} 
\int {\rm d}{\vec \phi} 
\ {\rm exp} \left\{ 
\frac{i \mu^{\prime}}{2} {\vec \phi}^2   
+ \frac{p}{N} \sum_{k=1}^M 
f({\vec \phi} \cdot {\vec \psi}_k) \right\} \right]^N 
\end{eqnarray}
and introducing 
\begin{equation}
{\tilde \eta}({\vec \psi})  = \frac{1}{M} \sum_{k=1}^M 
\delta({\vec \psi} - {\vec \psi}_k),
\end{equation}
we obtain 
\begin{eqnarray}
\langle Z^n \rangle  
& \sim & 
\int \prod_{k=1}^M {\rm d}{\vec \psi}_k 
\ {\rm exp}\left\{ M \frac{i}{2} 
\int {\rm d}{\vec \psi}\, {\tilde \eta}({\vec \psi}) {\vec \psi}^2 \right\} 
\nonumber \\ & \times & 
\left[ \frac{1}{(2 \pi i)^{\alpha n/2}} 
\int {\rm d}{\vec \phi} 
\ {\rm exp} \left\{ 
\frac{i \mu^{\prime}}{2} {\vec \phi}^2   
+ \alpha p \int {\rm d}{\vec \psi}\, {\tilde \eta}({\vec \psi}) 
f({\vec \psi} \cdot {\vec \phi}) \right\} \right]^N. \nonumber \\  
\end{eqnarray}
\par
As before, introducing an order parameter function $\eta({\vec \psi})$ satisfying 
the normalization 
\begin{equation}
\int \eta({\vec \psi})\, {\rm d}{\vec \psi} = 1,
\end{equation}
we can derive
\begin{equation}
\langle Z^n \rangle \sim \frac{1}{(2 \pi i)^{Mn/2}} \int {\cal D}\eta({\vec \psi}) \ {\rm e}^{N S[\eta]}.
\end{equation}
Here
\begin{eqnarray}
S[\eta] & = & - \alpha \int {\rm d}{\vec \psi}\, 
\eta({\vec \psi}) \ln \eta({\vec \psi}) 
+ \frac{i \alpha}{2} \int {\rm d}{\vec \psi}\, \eta({\vec \psi}) 
{\vec \psi}^2 \nonumber \\ 
& + & \ln \left[ \int {\rm d}{\vec \phi} 
\ {\rm exp}\left\{ \frac{i \mu^{\prime}}{2} {\vec \phi}^2 + \alpha p 
\int {\rm d}{\vec \psi}\, 
\eta({\vec \psi}) f({\vec \psi} \cdot {\vec \phi}) \right\} \right]. 
\end{eqnarray}
\par
In order to approximately evaluate the asymptotic behavior in  
the limit $N \rightarrow \infty$, we again introduce a Gaussian 
ansatz
\begin{equation}
\eta^{\rm DEMA}({\vec \psi}) = \frac{1}{(2 \pi i \tau)^{n/2}} \ {\rm exp}\left\{ 
- \frac{{\vec \psi}^2}{2 i \tau} \right\}, \ \ \ 
{\rm Im}\tau(\mu^{\prime}) \leq 0, 
\end{equation}
where $\tau$ is a parameter, 
and solve the stationary point equation  
\begin{equation}
\label{vdema}
\frac{\partial S[\eta^{\rm DEMA}]}{\partial \tau} = 0.  
\end{equation}
We shall call this scheme the Dual Effective Medium 
Approximation (DEMA). 
\par
Putting the Gaussian ansatz into (\ref{vdema}) 
and taking the limit $n \rightarrow 0$, we find
\begin{equation}
\alpha - \alpha \tau - 1 + \mu \sum_{k=0}^{\infty} 
\frac{{\rm e}^{-\alpha p} (\alpha p)^k}{k!} 
\int \prod_{j=1}^k \left( \Pi(x_j) {\rm d}x_j \right) 
\frac{1}{\mu - \tau \sum_{j=1}^k x_j^2} = 0.
\end{equation}
If $\tau$ satisfies this equation, we can write the 
spectral density as
\begin{eqnarray}
\rho(\mu) & = & \lim_{n \rightarrow 0}
\frac{2}{\pi n} {\rm Im}  
\frac{\partial}{\partial \mu} \frac{1}{N} \ln \left\langle 
(Z(\mu))^n \right\rangle  \sim \lim_{n \rightarrow 0}
\frac{2}{\pi n} {\rm Im}  
\frac{\partial}{\partial \mu} S[\eta^{\rm DEMA}] 
\nonumber \\ 
& = & - \frac{1}{\pi} {\rm Im} \left[ 
\sum_{k=0}^{\infty} 
\frac{{\rm e}^{-\alpha p} (\alpha p)^k}{k!} 
\int \prod_{j=1}^k \left( \Pi(x_j) {\rm d}x_j \right) 
\frac{1}{\mu - \tau \sum_{j=1}^k x_j^2} \right]
\nonumber \\ 
& = & - \frac{1}{\pi} {\rm Im} 
\frac{\alpha \tau - \alpha + 1}{\mu} = 
- \frac{\alpha}{\mu \pi} {\rm Im}\tau.     
\end{eqnarray}
\par
Let us suppose that we calculate the 
spectral density of ${\tilde J} = A A^{\rm T}$ 
as well. Although the exact density of the 
nonzero eigenvalues of ${\tilde J}$ must be 
identical to that of $J$, an approximate result 
can be different. In fact, there is a duality 
relation between EMA and DEMA: the EMA result for 
$J$ is the same as the DEMA result for ${\tilde J}$, 
and the DEMA result for $J$ is the same as 
the EMA result for ${\tilde J}$.   

\section{Symmetric Effective Medium Approximation}
\setcounter{equation}{0}
\renewcommand{\theequation}{4.\arabic{equation}}

The approximation schemes so far discussed need
numerical treatment of the stationary point equations. 
In this section, we introduce a symmetrized scheme of 
approximation, which we call the Symmetric 
Effective Medium Approximation (SEMA). In SEMA analytic 
solutions can be deduced, if the distribution 
of the nonzero elements of $A$ is binary. 
\par
In terms of the functions 
\begin{equation}
{\tilde \xi}({\vec \phi})  = \frac{1}{N} \sum_{j=1}^N 
\delta({\vec \phi} - {\vec \phi}_j), \ \ \ 
{\tilde \eta}({\vec \psi})  = \frac{1}{M} \sum_{k=1}^M 
\delta({\vec \psi} - {\vec \psi}_k),
\end{equation}
(\ref{zn1}) can be expressed in the form    
\begin{eqnarray}
\langle Z^n \rangle  
& \sim & 
\frac{1}{(2 \pi i)^{M n/2}} 
\int \prod_{j=1}^N {\rm d}{\vec \phi}_j 
\int \prod_{k=1}^M {\rm d}{\vec \psi}_k  
\nonumber \\ & \times & 
{\rm exp}\left\{ N \frac{i \mu^{\prime}}{2} 
\int {\rm d}{\vec \phi}\, {\tilde \xi}({\vec \phi}) 
{\vec \phi}^2 + M \frac{i}{2} 
\int {\rm d}{\vec \psi}\, {\tilde \eta}({\vec \psi}) 
{\vec \psi}^2 \right\} 
\nonumber \\ & \times & 
{\rm exp} \left\{ M p \int {\rm d}{\vec \psi} \int {\rm d}
{\vec \phi}\, {\tilde \eta}({\vec \psi}) {\tilde \xi}({\vec \phi}) 
f({\vec \psi} \cdot {\vec \phi}) \right\}.
\end{eqnarray}
As before it can be further rewritten as 
\begin{equation}
\langle Z^n \rangle \sim  
\frac{1}{(2 \pi i)^{M n/2}} 
\int {\cal D}\xi({\vec \phi}) {\cal D}\eta({\vec \psi}) 
\ {\rm e}^{ N S[\xi,\,\eta]},  
\end{equation}
where
\begin{eqnarray}
\label{ssema}
S[\xi,\,\eta] & = &  - \int {\rm d}{\vec \phi}\, \xi({\vec \phi}) 
\ln \xi({\vec \phi}) - \alpha \int {\rm d}{\vec \psi}\, \eta({\vec \psi}) 
\ln \eta({\vec \psi}) \nonumber \\ 
& + &  
\frac{i \mu^{\prime}}{2} 
\int {\rm d}{\vec \phi}\, \xi({\vec \phi}) 
{\vec \phi}^2 + \alpha \frac{i}{2} 
\int {\rm d}{\vec \psi}\, \eta({\vec \psi}) 
{\vec \psi}^2 \nonumber \\ 
& + &  
\alpha p \int {\rm d}{\vec \psi} \int {\rm d}
{\vec \phi}\, \eta({\vec \psi}) \xi({\vec \phi}) 
f({\vec \psi} \cdot {\vec \phi}).
\end{eqnarray}
It should be noted that one obtains $S[\xi]$ from $S[\xi,\,\eta]$ 
if one eliminates $\eta(\vec{\psi})$ from $S[\xi,\,\eta]$ using 
the stationary condition with respect to $\eta(\vec{\psi})$. 
Alternatively, eliminating $\xi(\vec{\phi})$ from $S[\xi,\,\eta]$ 
yields $S[\eta]$. 
\par
In SEMA we suppose that both of the approximate solutions 
for $\xi$ and $\eta$ have Gaussian 
forms 
\begin{equation}
\label{sigmatau}
\xi^{\rm SEMA}({\vec \phi}) = \frac{1}{(2 \pi i \sigma)^{n/2}} \ {\rm exp}\left\{ 
- \frac{{\vec \phi}^2}{2 i \sigma} \right\}, \ \ \ 
\eta^{\rm SEMA}({\vec \psi}) = \frac{1}{(2 \pi i \tau)^{n/2}} \ {\rm exp}\left\{ 
- \frac{{\vec \psi}^2}{2 i \tau} \right\} 
\end{equation}
(${\rm Im}\sigma(\mu^{\prime}) \leq 0$, ${\rm Im}\tau(\mu^{\prime}) \leq 0$). 
Solving the modified stationary point equations 
\begin{equation}
\frac{\partial S[\xi^{\rm SEMA},\,\eta^{\rm SEMA}]}{\partial \sigma} = 
\frac{\partial S[\xi^{\rm SEMA},\,\eta^{\rm SEMA}]}{\partial \tau} = 0,  
\end{equation}
we find the relations
\begin{eqnarray}
\label{sema}
1 - \sigma \mu + \alpha p \sigma \tau \int {\rm d}x\, \Pi(x) 
\frac{x^2}{1 - \sigma \tau x^2}  
& = & 0, \nonumber \\ 
\alpha - \alpha \tau + \alpha p \sigma \tau \int {\rm d}x\, \Pi(x) 
\frac{x^2}{1 - \sigma \tau x^2}  
& = & 0,
\end{eqnarray}
from which it follows that
\begin{equation}
\tau = \frac{\mu \sigma + \alpha - 1}{\alpha}.
\end{equation}
Using the solution $\sigma$ of (\ref{sema}), as before 
we obtain the spectral density as  
\begin{equation}
\rho(\mu) \sim  \lim_{n \rightarrow 0}
\frac{1}{\pi n} {\rm Re} \int {\rm d}{\vec \phi}\,\xi({\vec \phi}) 
{\vec \phi}^2 = - \frac{1}{\pi} {\rm Im}\sigma.   
\end{equation}

\section{Binary Distribution}
\setcounter{equation}{0}
\renewcommand{\theequation}{5.\arabic{equation}}

In this section, we calculate approximate spectral densities 
of the matrix $J = A^{\rm T} A$, employing EMA and 
SEMA as the approximation schemes. For the nonzero elements 
of the $M \times N$ real matrix $A$, we assume in this section the 
binary distribution
\begin{equation}
\label{binary}
\Pi(x) = \frac{1}{2} \left( \delta(x - 1) + \delta(x + 1) \right).
\end{equation} 
In this typical case, the Fourier transform of $P(x)$ 
is given by 
\begin{equation}
{\hat P}(k) =  1 + \frac{p}{N} (\cos k - 1).  
\end{equation}
Namely, we have 
\begin{equation}
f(k) =  \cos k - 1.  
\end{equation}

\par
For the binary distribution (\ref{binary}), the stationary 
point equation (\ref{ema}) for EMA takes the form
\begin{equation}
1 - \sigma \mu - \alpha + \alpha \sum_{k=0}^{\infty} 
\frac{{\rm e}^{-p} p^k}{k!} \frac{1}{1 - \sigma k} = 0.
\end{equation}
As we mentioned, we are able to numerically solve 
this equation. 
\par
Comparison of the EMA solution and the average spectral density 
of numerically generated random matrices (NUMERICAL) is shown 
in Figure 1. One can see that the agreement 
is reasonably good except in the tail region.

\begin{figure}[h]
\epsfxsize=11cm
\centerline{\epsfbox{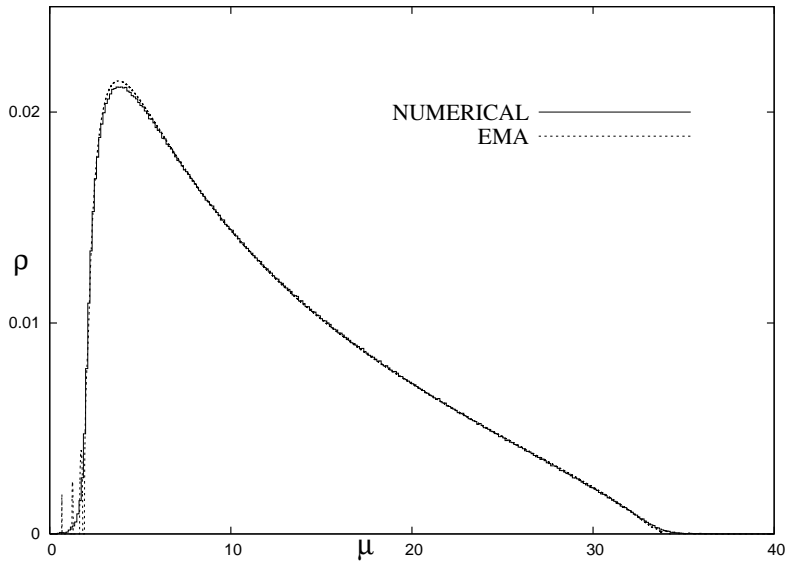}}
\caption{Comparison of the EMA solution ($p = 12, \alpha = 0.3$) 
for the spectral density and a numerically generated result 
(average over $1000$ samples, $N = 8000$).}
\end{figure}

\par 
On the other hand, for the binary distribution (\ref{binary}), 
the SEMA stationary point equations (\ref{sema}) are written as  
\begin{eqnarray}
1 - \sigma \mu + \alpha p \frac{\sigma \tau}{1 - \sigma \tau} 
& = & 0, \nonumber \\ 
\alpha - \alpha \tau + \alpha p \frac{\sigma \tau}{1 - \sigma \tau} 
& = & 0.
\end{eqnarray}
Then it is straightforward to derive a cubic equation for $\sigma$
\begin{equation}
\sigma^3 + \frac{\alpha (p + 1) - 2}{\mu} \sigma^2 - 
\frac{\mu \alpha + (1 - \alpha)(\alpha p - 1)}{\mu^2} \sigma + 
\frac{\alpha}{\mu^2} = 0.
\end{equation}
This cubic equation can be analytically solved. Using the notations 
\begin{equation}
a_2 =  \frac{\alpha (p + 1) - 2}{\mu}, \ \ \  
a_1 =  - \frac{\mu \alpha + (1 - \alpha)(\alpha p - 1)}{\mu^2}, \ \ \  
a_0 = \frac{\alpha}{\mu^2}
\end{equation}
for the coefficients, we define
\begin{equation}
q = \frac{1}{3} a_1 - \frac{1}{9} a_2^2,  \ \ \ 
r = \frac{1}{6} (a_1 a_2 - 3 a_0) - \frac{1}{27} a_2^3.
\end{equation}
Depending on the value of $q^3 + r^2$, the solutions 
are classified as
\par
\medskip
1. If $q^3 + r^2 > 0$, there are one real and two complex solutions.
\par
\medskip
2. If $q^3 + r^2 = 0$, there are three real solutions (at least two 
of them are the same).
\par
\medskip
3. If $q^3 + r^2 < 0$, there are three real solutions. 
\par
\medskip
In order to have a nonzero eigenvalue density, we need 
to have a complex solution with a negative imaginary part. 
Therefore we focus on the case $q^3 + r^2 > 0$. 
For a complex number  $z = |z| \ {\rm e}^{i \theta}$ 
($- \pi < \theta \leq \pi$) and 
a real number $\chi$, let us define the 
exponential function as 
\begin{equation}
z^{\chi} = |z|^{\chi} \ {\rm e}^{i \chi \theta}.
\end{equation}
Then, using real numbers 
\begin{equation}
s_1 = - (- r - (q^3 + r^2)^{1/2})^{1/3}, \ \ \ 
s_2 = - (- r + (q^3 + r^2)^{1/2})^{1/3}, 
\end{equation}
one obtains a real solution
\begin{equation}
z_0 = s_1 + s_2 - \frac{a_2}{3}
\end{equation}
and complex solutions
\begin{eqnarray}
\label{z1z2}
z_1 & = & - \frac{1}{2} (s_1 + s_2) - \frac{a_2}{3} 
+ i \frac{\sqrt{3}}{2} (s_1 - s_2), \nonumber \\ 
z_2 & = & - \frac{1}{2} (s_1 + s_2) - \frac{a_2}{3} 
-  i \frac{\sqrt{3}}{2} (s_1 - s_2).
\end{eqnarray}
The nonzero eigenvalue density is thus given by
\begin{eqnarray}
\label{rhosema}
\rho(\mu) = - \frac{1}{\pi} {\rm Im}z_2 = 
\frac{\sqrt{3}}{2 \pi} (s_1 - s_2).  
\end{eqnarray}
This equation gives a continuous band between the edges 
$\mu_-$ and $\mu_+$ ($\mu_- < \mu_+$), where $q^3 + r^2 = 0$ 
holds.

\begin{figure}[h]
\epsfxsize=11cm
\centerline{\epsfbox{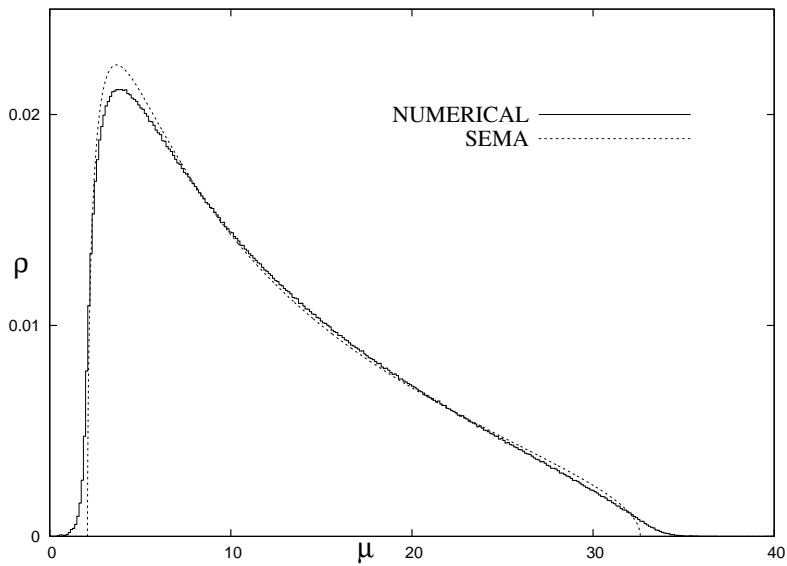}}
\caption{Comparison of the SEMA solution ($p = 12, \alpha = 0.3$) 
for the spectral density and a numerically generated 
result (average over $1000$ samples, $N = 8000$).}
\end{figure}
 
\par
We compare the SEMA solution (\ref{rhosema}) 
and a numerically generated result (NUMERICAL) 
in Figure 2. In spite of the simplification of the 
scheme, the agreement is still reasonably good. 
However, as the SEMA spectral density vanishes out of the 
edges $\mu_-$ and $\mu_+$, the discrepancy in the 
tail region stands out. 

\begin{figure}[h]
\epsfxsize=11cm
\centerline{\epsfbox{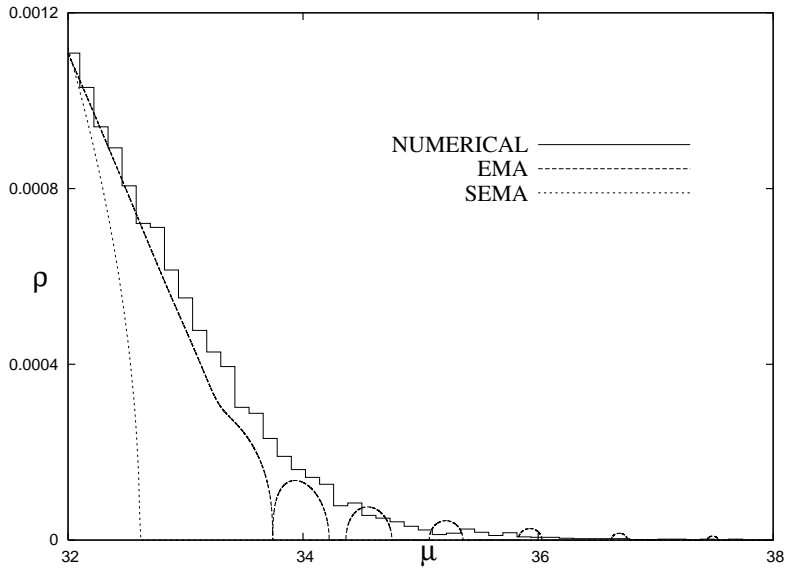}}
\caption{Comparison of the EMA and SEMA solutions 
($p = 12, \alpha = 0.3$) and a numerically generated 
result (average over $1000$ samples, $N = 8000$) 
in the neighborhood of the outer edge $\mu_+ = 32.611$.}
\end{figure}

\par
In Figure 3, we compare the EMA and SEMA solutions and 
a numerically generated result (NUMERICAL) in the 
neighborhood of the outer edge $\mu_+$. One can see that 
the EMA solution has minibands out of the outer edge, 
which approximate the tail spectrum of the numerical 
result. On the other hand, the neighborhood of the inner 
edge $\mu_-$ is depicted in Figure 4. The EMA 
solution also has minibands within the gap 
between the origin and the inner edge. 

\begin{figure}[h]
\epsfxsize=11cm
\centerline{\epsfbox{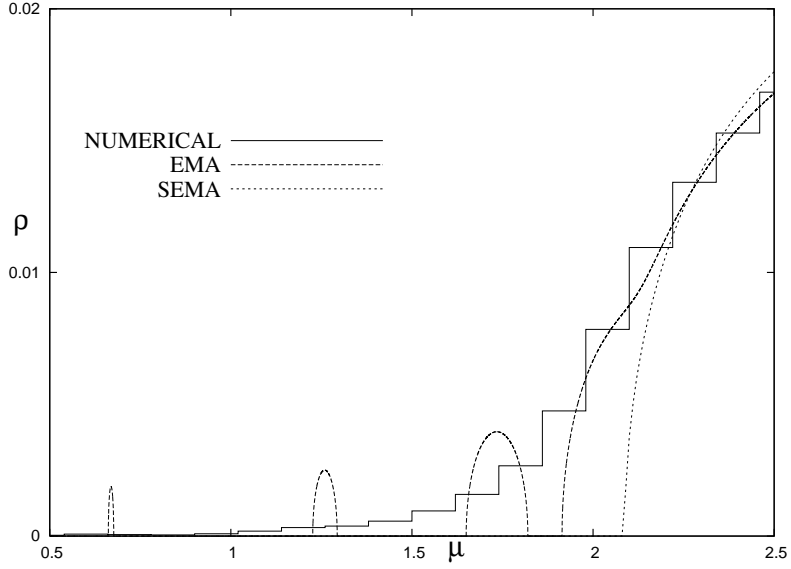}}
\caption{Comparison of the EMA and SEMA solutions 
($p = 12, \alpha = 0.3$) and a numerically generated 
result (average over $1000$ samples, $N = 8000$) 
in the neighborhood of the inner edge $\mu_- = 2.087$.}
\end{figure}

\section{Single Defect Approximation}
\setcounter{equation}{0}
\renewcommand{\theequation}{6.\arabic{equation}}

In SEMA, the spectral density vanishes out of 
the edges $\mu_-$ and $\mu_+$. Therefore, in order 
to analyze the tail region, we need to improve the 
approximation. Let us begin with the definition 
(\ref{ssema}) of $S[\xi,\,\eta]$. The exact stationary 
point equation can be derived from the variational 
equation
\begin{equation}
\delta \left\{ S[\xi,\,\eta] 
+ a \left( \int {\rm d}{\vec \phi}\, 
\xi({\vec \phi}) - 1 \right)     
+ b \left( \int {\rm d}{\vec \psi}\, 
\eta({\vec \psi})- 1  \right) \right\}  
= 0
\end{equation}
as    
\begin{eqnarray}
\frac{\delta S[\xi,\,\eta]}{\delta \xi({\vec \phi})} + a  & = &  
- \ln \xi({\vec \phi}) - 1 + \frac{i \mu^{\prime}}{2} {\vec \phi}^2 
+ \alpha p \int {\rm d}{\vec \psi}\, \eta({\vec \psi}) 
f({\vec \psi} \cdot {\vec \phi}) + a = 0, \nonumber \\ 
\frac{\delta S[\xi,\,\eta]}{\delta \eta({\vec \psi})} + b  & = &  
- \alpha \ln \eta({\vec \psi}) - \alpha + \frac{i \alpha}{2} {\vec \psi}^2 
+ \alpha p \int {\rm d}{\vec \phi}\, \xi({\vec \phi}) 
f({\vec \psi} \cdot {\vec \phi}) + b = 0. \nonumber \\ 
\end{eqnarray} 
Here $a$ and $b$ are the Lagrange multipliers which ensure 
the normalizations of $\xi({\vec \phi})$ and $\eta({\vec \psi})$. 
Then we find
\begin{eqnarray}
\label{xi}
\xi({\vec \phi}) & = & {\cal A} \ {\rm exp}\left[ \frac{i \mu^{\prime}}{2} {\vec \phi}^2 
+ \alpha p \int {\rm d}{\vec \psi}\, \eta({\vec \psi}) f({\vec \psi} \cdot 
{\vec \phi}) \right], \\ 
\label{eta}
\eta({\vec \psi}) & = & {\cal B} \ {\rm exp}\left[ \frac{i}{2} {\vec \psi}^2 
+ p \int {\rm d}{\vec \phi}\, \xi({\vec \phi}) f({\vec \psi} \cdot 
{\vec \phi}) \right],
\end{eqnarray}
where ${\cal A}$ and ${\cal B}$ should be determined so that 
$\xi({\vec \phi})$ and $\eta({\vec \psi})$ are 
correctly normalized.
\par 
We have so far made no approximation. Now, as the first improvement 
of SEMA, we put the Gaussian ansatz   
\begin{equation}
\eta({\vec \psi}) = \frac{1}{(2 \pi i \tau)^{n/2}} \ {\rm exp}\left\{ 
- \frac{{\vec \psi}^2}{2 i \tau} \right\}, \ \ \ 
{\rm Im}\tau(\mu^{\prime}) \leq 0
\end{equation}
into the RHS of (\ref{xi}) and calculate the LHS. Then, assuming that 
$\tau$ is the solution of SEMA, we obtain the improved approximation 
for $\xi({\vec \phi})$ as the LHS. This approximation scheme 
is called the single defect approximation (SDA)~\cite{SC,BM}. 
\par   
For the binary distribution (\ref{binary}), it follows from  
$f(k) = \cos k - 1$ that
\begin{eqnarray}
\xi({\vec \phi}) & = & {\cal A} \ {\rm e}^{i \mu^{\prime} {\vec \phi}^2/2} 
\sum_{k=0}^{\infty} \frac{{\rm e}^{- \alpha p} (\alpha p)^k}{k!} 
\left[ \int {\rm d}{\vec \psi}\, \eta({\vec \psi}) \cos({\vec \psi} 
\cdot {\vec \phi}) \right]^k  \nonumber \\ 
& = & {\cal A} 
\sum_{k=0}^{\infty} \frac{{\rm e}^{- \alpha p} (\alpha p)^k}{k!} 
\ {\rm exp}\left[ \frac{i}{2} (\mu^{\prime} - k \tau) {\vec \phi}^2 \right].
\end{eqnarray}
Now we are able to determine the normalization constant ${\cal A}$. 
Integrating the both sides of the above equation over 
${\vec \phi}$, we find 
\begin{equation}
\int \xi({\vec \phi})\,{\rm d}{\vec \phi} =  
{\cal A} \sum_{k=0}^{\infty} \frac{{\rm e}^{- \alpha p} (\alpha p)^k}{k!} 
\left[ \frac{i}{2 \pi} (k \tau - \mu^{\prime}) \right]^{-n/2}.
\end{equation}
Hence ${\cal A}$ should be set to $1$ in the limit $n \rightarrow 0$. 
Then the spectral density can be evaluated as
\begin{eqnarray}
\label{eq:sdSEMA}
\rho(\mu) & = &  \lim_{n \rightarrow 0} \frac{1}{\pi n} 
{\rm Re}\int {\rm d}{\vec \phi}\, \xi({\vec \phi}) {\vec \phi}^2 
\nonumber \\ & = &  \lim_{n \rightarrow 0} \frac{1}{\pi n} 
{\rm Re} \sum_{k=0}^{\infty} \frac{{\rm e}^{- \alpha p} (\alpha p)^k}{k!} 
\int {\rm d}{\vec \phi}\, {\vec \phi}^2 
\ {\rm exp}\left[ \frac{i}{2} (\mu^{\prime} - k \tau) {\vec \phi}^2 \right]
\nonumber \\ 
&  = &  - \frac{1}{\pi} 
{\rm Im} \sum_{k=0}^{\infty} \frac{{\rm e}^{- \alpha p} 
(\alpha p)^k}{k!} \frac{1}{\mu^{\prime} - k \tau}.
\end{eqnarray}
Out of the edges $\mu_-$ and $\mu_+$, the SEMA solution 
$\tau(\mu)$ is real. Therefore the spectral density~(\ref{eq:sdSEMA}) has 
delta peaks 
\begin{equation}
\label{delta}
\frac{{\rm e}^{- \alpha p} (\alpha p)^k}{k!} 
\frac{1}{1 - k \tau'(\mu_k)} \delta(\mu - \mu_k),
\end{equation}
where $\mu_k$ satisfies the equation  
\begin{equation}
\mu_k - k \tau(\mu_k) = 0.
\end{equation}
\par
Let us note that the imaginary part of $\sigma(\mu + i \epsilon)$ 
and $\tau(\mu + i \epsilon)$ cannot be positive (see 
(\ref{sigmatau})). This means that 
\begin{equation}
{\rm Im}\sigma(\mu) < 0
\end{equation}
or, if ${\rm Im}\sigma(\mu) = 0$, 
\begin{equation}
\label{condition}
{\rm Re} \frac{{\rm d}\sigma}{{\rm d} \mu} \leq 0, \ \ \  
{\rm Re} \frac{{\rm d}\tau}{{\rm d}\mu} = \frac{1}{\alpha} {\rm Re} 
\frac{{\rm d}(\mu \sigma)}{{\rm d}\mu} \leq 0. 
\end{equation}
\par
When $\mu$ is larger than the outer edge $\mu_+$ of 
the band, the solution $z_1$ satisfies these conditions. 
The locations of the delta peaks are thus determined 
by solving the equation  
\begin{equation}
\mu - k \frac{\mu z_1(\mu) + \alpha - 1}{\alpha} = 0, \ \ \ 
\mu > \mu_+.
\end{equation}
For example, if $p = 12$ and $\alpha = 0.3$, the locations 
of the delta peaks are
\begin{equation}
\mu = 32.667, \ 32.926, \ 33.336, \ 33.856, \ 34.456, \ 35.118, 
\ 35.828,\cdots,
\end{equation}
corresponding to $k = 14,15,16,17,18,19,20,\cdots$ 
($\mu_+ = 32.611$). We would like to note that the 
DEMA minibands (not shown) seem to appear around these delta 
peaks. 
\par
From the analytic solution (\ref{z1z2}), we find the asymptotics
\begin{equation}
z_1(\mu) \sim \frac{1}{\mu}, \ \ \ \mu \rightarrow \infty,
\end{equation}
from which it follows that
\begin{equation}
\tau(\mu) = \frac{\mu z_1(\mu) + \alpha - 1}{\alpha} \sim 1, \ \ \ 
\mu \rightarrow \infty. 
\end{equation}
Thus the delta peaks are located at $\mu_k \sim k$, 
so that a continuous approximation gives 
\begin{equation}
\rho(\mu_k) (\mu_k - \mu_{k-1}) \sim  
\frac{{\rm e}^{- \alpha p} (\alpha p)^k}{k!}. 
\end{equation}
Therefore, in the limit of large $\mu$, we obtain
\begin{equation}
\rho(\mu) \sim \frac{{\rm e}^{-\alpha p}}{\sqrt{2 \pi \mu}} 
\ {\rm exp}\left[ - \mu \ln \left( \frac{\mu}{\alpha p {\rm e}} \right) \right].
\end{equation}
A similar asymptotic formula is known
for real symmetric sparse random matrices $R$ whose elements 
$R_{jl}$ ($j \leq l$) are independently distributed~\cite{RB,SC}. 
As argued in \cite{SC}, SDA is expected to be correct for large $\mu$. 
However, the weights of the delta peaks (\ref{delta}) are too small to 
account for the numerically calculated cumulative density. In order to 
improve the agreement, we presumably need to develop higher order 
extensions of SDA by iterating the approximation scheme.
\par
When $\mu$ is smaller than the inner edge $\mu_-$, the solution 
$z_2$ satisfies the conditions (\ref{condition}). However, 
as $\mu z_2 + \alpha - 1 \leq 0$, there is no SDA delta peak 
in this region besides at the origin. Therefore the second order 
extension of SDA should be applied. Assuming that $\sigma$ is the 
solution $z_2$ of SEMA, we put the Gaussian 
ansatz   
\begin{equation}
\xi({\vec \phi}) = \frac{1}{(2 \pi i \sigma)^{n/2}} \ {\rm exp}\left\{ 
- \frac{{\vec \phi}^2}{2 i \sigma} \right\} 
\end{equation}
into the RHS of (\ref{eta}). Then $\eta$ is calculated as 
the LHS of (\ref{eta}) and put into the RHS of (\ref{xi}). 
Thus the LHS of (\ref{xi}) gives the approximate solution 
for $\xi({\vec \phi})$. Within this approximation, an argument 
as before yields an equation  
\begin{equation}
\mu - \sum_{k=0}^{\infty} \frac{n_k}{1 - k z_2(\mu)} 
= 0, \ \ \  
\mu < \mu_-,
\end{equation}
which determines the locations of the delta peaks. 
Here $n_k$ are non-negative integers. For example, 
if $p = 12$ and $\alpha = 0.3$,  we find the delta 
peaks at $\mu = 1,\ 1.6,\ 2,\ 2.035$ besides at the 
origin ($\mu_- = 2.087$).

\section{Summary}
\setcounter{equation}{0}
\renewcommand{\theequation}{7.\arabic{equation}}

In this paper, we evaluated the asymptotic eigenvalue density 
of the matrix of the form $J = A^{\rm T} A$, where $A$ is an 
$M \times N$ real sparse random matrix. We utilized replica 
method and developed approximation schemes 
called the Effective Medium Approximation (EMA) and the Dual Effective 
Medium Approximation (DEMA). Moreover a symmetrized version of 
EMA (SEMA) was presented. In the case of binary distribution 
of the nonzero elements of $A$, analytic solutions were 
derived for SEMA. We compared the results of EMA and SEMA, 
focusing on the behavior in the tail region. In order to 
analytically improve SEMA, we further developed the Single Defect 
Approximation (SDA) and evaluated the tail behavior. 

\section*{Acknowledgements}

One of the authors (T.N.) is grateful to Dr.\ Laurent Laloux, 
Dr.\ Marc Potters and Prof.\ Jean-Philippe Bouchaud for suggesting 
the subject of this article. He also thanks Prof.\ Geoff J. Rodgers 
for valuable discussions. The authors acknowledge support 
from the Grant-in-Aid for Scientific Research, MEXT, 
Japan (Nos.~16740224, 14084209 and 18079010).


\begin{thebibliography}{notitle}

\bibitem{JW}
J. Wishart, Biometrika {\bf 20} (1928) 32.  
\bibitem{CKS}
Y. Le Cun, I. Kanter and S.A. Solla, Phys. Rev. Lett. {\bf 66} (1991) 2396.
\bibitem{JV}
J.J.M. Verbaarschot, Acta Phys. Polon. {\bf B25} (1994) 133.
\bibitem{CB}
C.W.J. Beenakker, Rev. Mod. Phys. {\bf 69} (1997) 731. 
\bibitem{LCBP}
L. Laloux, P. Cizeau, J.-P. Bouchaud and M. Potters, Phys. Rev. Lett. 
{\bf 83} (1999) 1467.
\bibitem{PLEROU}
V. Plerou, P. Gopikrishnan, B. Rosenow, L.A. Nunes Amaral and H.E. Stanley, 
Phys. Rev. Lett. {\bf 83} (1999) 1471.
\bibitem{TV}
A.M. Tulino and S. Verd\'u, Foundations and Trends in Communications 
and Information Theory {\bf 1} (2004) Issue 1. 
\bibitem{WIG}
E.P. Wigner, Ann. Math. {\bf 67} (1958) 325.  
\bibitem{MP}
V. Mar\u{c}enko and L. Pastur, Math. USSR-Sbornik {\bf 1} (1967) 457.
\bibitem{PASTUR}
L. Pastur, {\it Mathematical Physics 2000} (Imperial College Press, 2000) 216. 
\bibitem{YT}
M. Yoshida and T. Tanaka, ``Analysis of Sparsely-Spread CDMA via Statistical
Mechanics'', {\em Proc. 2006 IEEE Int. Symp. Info. Theory} (2006) pp. 2378--2382.
\bibitem{SOSH}
A. Soshnikov, Elect. Comm. in Probab. {\bf 9} (2004) 82. 
\bibitem{SF}
A. Soshnikov and Y.V. Fyodorov, J. Math. Phys. {\bf 46} (2005) 033302.
\bibitem{BBP}
G. Biroli, J.-P. Bouchaud and M. Potters, preprint(cond-mat/0609070).
\bibitem{TT}
T. Tanaka, ``On the Eigenvalue Spectrum of Random Matrices'', 
extended abstract, {\it Randomness and Computation Joint Workshop 
``New Horizons in Computing'' and ``Statistical Mechanical Approach 
to Probabilistic Information Processing'' (18-21 July, 2005, Sendai, 
Japan)}.
\bibitem{RB}
G.J. Rodgers and A.J. Bray, Phys. Rev. {\bf B37} (1988) 3557. 
\bibitem{RD}
G.J. Rodgers and C. De Dominicis, J. Phys. {\bf A23} (1990) 1567. 
\bibitem{MF}
A.D. Mirlin and Y.V. Fyodorov, J. Phys. {\bf A24} (1991) 2273. 
\bibitem{SC}
G. Semerjian and L.F. Cugliandolo, J. Phys. {\bf A35} (2002) 4837. 
\bibitem{BM}
G. Biroli and R. Monasson, J. Phys. {\bf A32} (1999) L255. 

\end{thebibliography}
\end{document}